\documentstyle[astron]{l-aa}
\begin{document}
\input{psfig}
\thesaurus{}
\title{The distance of the Fornax Cluster based on Globular Cluster
Luminosity Functions}

\author{Sven Kohle \inst{1}
\and  Markus Kissler-Patig \inst{1,2}
\and  Michael Hilker \inst{1}
\and  Tom Richtler \inst{1,3}
\and  L. Infante \inst{3}
\and  H. Quintana \inst{3}}

\offprints{S. Kohle, Bonn (skohle@astro.uni-bonn.de)}

\institute{
Sternwarte der Universit\"at Bonn, Auf dem H\"ugel 71, D-53121 Bonn, 
F.R. Germany
\and 
European Southern Observatory, Karl-Schwarzschildstr. 2, D-85748 Garching,
\and 
Department of Astronomy and Astrophysics, P. Universidad Cat\'{o}lica, 
Casilla 104, Santiago 22, Chile}

\date{Received ; accepted }
\thesaurus{11(10.07.2 ; 11.03.4 ; 11.05.1 ; 11.09.1)}
\maketitle

\begin{abstract}

We present Globular Cluster Luminosity Functions for four ellipticals and
one S0-Galaxy in the Fornax cluster of galaxies, derived from CCD photometry 
in V and I. 
The averaged turnover magnitudes are \mbox{$V_{TO} = 23.80 \pm 0.06$ mag} 
and \mbox{$I_{TO} =22.39 \pm 0.05$ mag}, 
respectively. We derive a  relative distance modulus 
\mbox{$(m-M)_{Fornax} - (m-M)_{M87} = 0.08 \pm 0.09$ mag} using the 
turnover of M87 based on HST data.


\end{abstract}

\section{Introduction}

There is a tremendously rich literature on the distance determination of
the Virgo cluster (e.g. de Vaucouleur 1993\nocite{deVaucouleurs:1993}). 
Relatively little has been
worked on the Fornax cluster of galaxies, although it offers several
advantages with respect to the Virgo cluster: it is more compact, does
not show substructure, and is a spiral-poor, evolved galaxy cluster, where
one can suspect that the elliptical galaxies projected near the center are
indeed spatially concentrated.

In this Letter, we apply the method of globular cluster luminosity functions
(GCLFs) to derive the distances of four ellipticals in the Fornax cluster.
 The central point of this method, namely the universality of the turnover (TO)
magnitude is still a subject of some debate (e.g. Harris in Jacoby et al. 1993
\nocite{Jacoby:1992}, Sandage \& Tammann 1995, Richtler 1995). 
We show that among our target galaxies, the turnover magnitude scatters with a 
dispersion of 0.23 mag in V and only 0.14 mag in I, which indeed is evidence 
for a universal turnover.

\section{Data, reduction, photometric calibration}
The full presentation of the data, together with a general presentation of
the globular cluster system (GCS) will be given in a forthcoming paper, so
here we restrict ourselves to some basic remarks. 

Our targets were the galaxies NGC 1399, 1374, 1379, 1387, 1427 (Table 
\ref{observations}).  The observations
have been performed with the 2.5m DuPont telescope at Las Campanas Observatory,
Chile, in the period Sept. 26-29, 1994. The CCD was a Tektronix chip
with a pixel size of 0.227 \arcsec, resulting in a frame with dimension
$7.4 \arcmin \times 7.4 \arcmin$. Deep exposures have been gained in Johnson V 
and Cousins I. Only NGC 1399 was covered by four overlapping frames, 
the other targets are much less extended and fit each on one frame. 
The seeing conditions were of medium quality, ranging from 1.0\arcsec  
to 1.5\arcsec. The limiting magnitude was approximately V=24 mag.

Modelling and subtraction of the target galaxies has been done with 
IRAF routines, finding and photometry of point sources with DAOPHOT II 
under IRAF.
The differences between PSF and aperture photometry could be determined with
an accuracy of better than 0.015 mag.
The photometric calibration has also been done with IRAF routines, using
 standard stars from Landolt (1992)\nocite{Landolt:1992}.
About 30 standard star measurements were made nightly under photometric 
conditions.
The mean residual of calibrated magnitude to standard star magnitude is 
about 0.02 mag for each night.

\section{Extraction of the luminosity functions: Treatment of background
and completeness}

\begin{figure}[t]
\psfig{figure=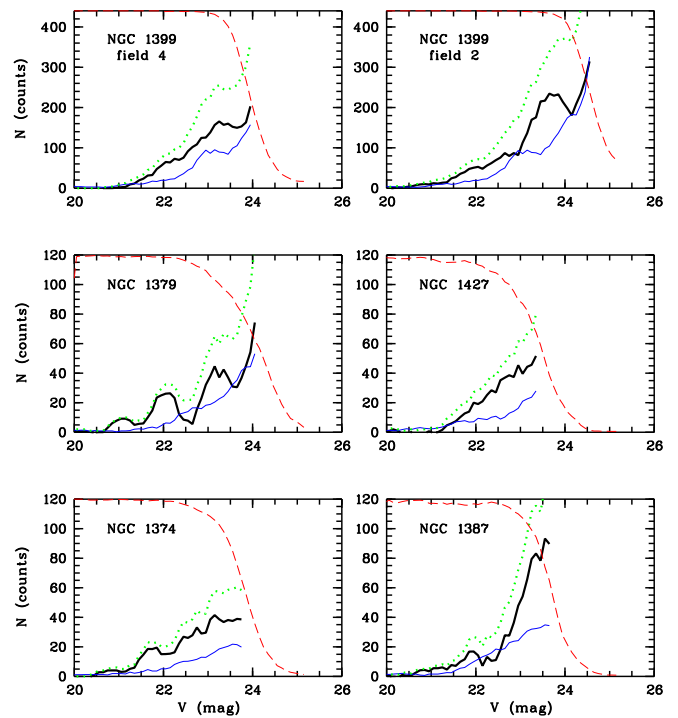,height=10cm,width=9cm}
\caption[]{This plot shows for our 5 target galaxies the completeness corrected
total counts (thick), background counts (dotted), the background subtracted counts
(thin), and the
completeness (dashed), binned in 0.5 mag intervals in V.  Obvious galaxies are already
removed.}
\label{background}
\end{figure}

Obviously, a proper knowledge of the background population is essential for 
deriving a GCLF, since we cannot distinguish between point-like background 
sources and globular clusters.
With the exception of NGC 1399, the GCSs of the galaxies are
considerably smaller in extent than the framesize so that the background could
be determined locally. A coarse selection of GC candidates with respect to 
extended objects was done by selecting sources with PSF profiles. The remaining
background population consists mostly of distant galaxies, to some extend
also of foreground stars. Leaving details for a forthcoming 
full data presentation, we show in Fig. \ref{background} for V the relative 
contribution of the total counts, the background, and the subtracted counts 
in dependence on magnitude. The dashed line shows in addition the completeness, 
calculated by artificial star experiments.\\

 \begin{table}[tbp]
 \caption[]{The observations of the Fornax galaxies obtained with the 2.5m 
telescope on Las Campanas}
     \begin{tabular}{ccccc}
       \hline
    galaxy & Filter & Obs. date & exposure & FWHM \\
    \hline
    NGC 1374 & V & 28 9 94 & $2 \times 1200$s & 1\farcs 2 \\
             & I & 29 9 94 & $1200$s, & 1\farcs 5 \\
             &   &         & $2\times 600$s & \\
    NGC 1379 & V & 29 9 94 & $2\times 1200$s & 1\farcs 2 \\
             & I & 26 9 94 & $3\times 1200$s & 1\farcs 4 \\
    NGC 1387 & V & 28 9 94 & $2\times 900$s & 1\farcs 3 \\
             & I & 28 9 94 & $2\times 900$s & 1\farcs 2 \\
    NGC 1427 & V & 26 9 94 & $3\times 1200$s & 1\farcs 5 \\
             & I & 26 9 94 & $3\times 1200$s & 1\farcs 3 \\
    NGC 1399 & I & 27/28 9 94 & $4\times 900$s & 1\farcs 2\\
             & V & 27/28 9 94 & $4\times 900$s & 1\farcs 0\\
    \hline
    \end{tabular}
 \label{observations}
  \end{table}

   \begin{table}[tb]
          
         \caption[]{Turnover values of the Fornax GCLFs fitted by 
gaussians and $t_{5}$ functions.}
             \begin{tabular}{ccc}
             \hline
             galaxy & $V^{Gauss}_{0}$ & $I^{Gauss}_{0}$ \\
             \hline
             NGC 1374 & 23.52 $\pm$ 0.14 & 22.60 $\pm$ 0.13 \\
             NGC 1379 & 23.68 $\pm$ 0.28 & 22.54 $\pm$ 0.34 \\
             NGC 1427 & 23.78 $\pm$ 0.21 & 22.31 $\pm$ 0.14 \\
             NGC 1399 & 23.90 $\pm$ 0.08 & 22.36 $\pm$ 0.06 \\
             \hline
           \end{tabular}

           \begin{tabular}{ccc}
             galaxy & $V^{t_{5}}_{0}$ & $I^{t_{5}}_{0}$ \\
             \hline
             NGC 1374 & 23.44 $\pm$ 0.13 & 22.49 $\pm$ 0.12 \\
             NGC 1379 & 23.65 $\pm$ 0.26 & 22.77 $\pm$ 0.44 \\
             NGC 1427 & 23.63 $\pm$ 0.19 & 22.22 $\pm$ 0.13 \\
             NGC 1399 & 23.90 $\pm$ 0.08 & 22.27 $\pm$ 0.05 \\
             \hline
            \end{tabular}
         \label{gauss}
   \end{table}

\section{Luminosity Functions}
During the last years, a lot of work has been done as for the correct 
representation of the GCLF (e.g. Secker et al. 1992\nocite{Secker:1992}). 
We cannot enter this discussion here, but give GCLFs for
two different representations with reference to the Milky Way system (see
next section): a Gaussian function with a fixed dispersion of 1.2
mag in V and and a $t_{5}$-function with a fixed dispersion of 1.1 mag. The counts
have been binned in 0.5 mag bins. The dispersions were determined by fitting GCLFs 
with free parameters but it should be emphasized that the dispersion must be 
fixed when comparing the TO of the Fornax GCLFs. 

\begin{figure*}[t]
\psfig{figure=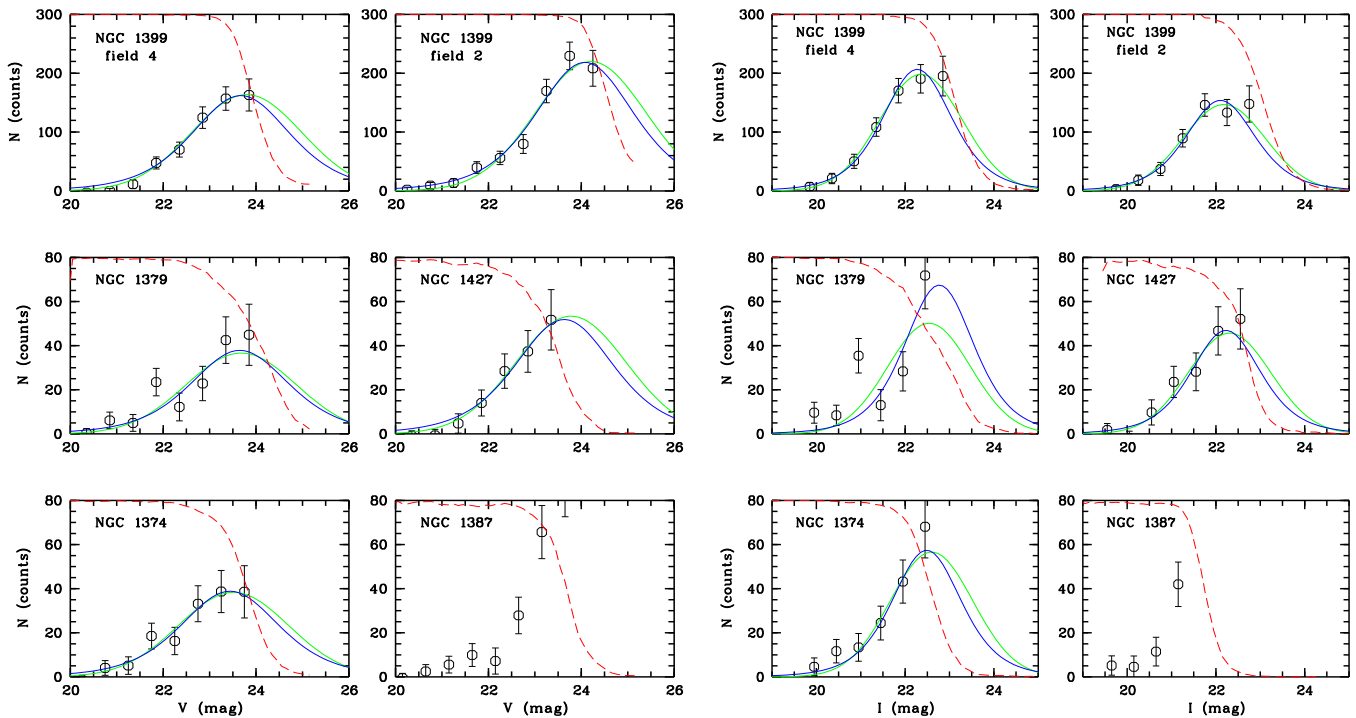,height=10cm,width=18.8cm}
\caption[]{Globular Cluster Luminosity Functions for the five Fornax galaxies in V an I. 
The GCLFs are corrected for completeness and background contamination. Overplotted are 
fitted gaussians (thick lines) and t${_5}$-functions (thin lines). The dashed line shows the 
completeness function calculated by artificial star experiments. NGC 1387 has a too sparse GCS
in order to fit a LF.}
\label{LFV} 
\end{figure*}

 Fig. \ref{LFV}  shows the results for the GCLFs in V and I, respectively.
Table \ref{gauss} summarises the results for different fits, done with 
Gaussians and $t_{5}$-functions.   
Because of its poor GCS  no GCLF of the S0 galaxy NGC 1387 could be fitted.
Taking averages for the turnover magnitudes, weighted with inverse 
square of the errors, we get for the mean turnover of the Fornax cluster 
$23.80 \pm 0.06$ in V, \mbox{$22.39 \pm 0.05$} in I for gaussians 
and  $23.73 \pm 0.07$ in V, $22.30 \pm 0.04$ in I for $t_{5}$. 
The given uncertainties represent the errors of the mean.

\section{The zero-point from the galactic globular cluster system}

The zero-point of the distance determination via GCLFs is given by the
galactic GCS. The uncertainty concerning distances
to galactic globular clusters still determines the absolute uncertainty
inherent to this method of deriving the distance of the Fornax cluster. For
the absolute turnover magnitude of the Milky Way system, one can
find values as different as 0.4 mag in the literature (e.g. Secker 1992, Sandage \&
Tamman 1995), related to the
adopted dependence of the brightness of horizontal branch stars on metallicity (see equations
below).
 This uncertainty will probably be overcome, once it is possible to derive GC distances from nearby
subdwarfs.
At present, one has to accept it and we give
the turnover magnitudes for two different relationships regarding the
metallicities of horizontal branch stars according to Sandage \& Tammann (1995)
 \nocite{Sandage:1995} and Harris et al. (1991) (note that Carney et al. 1992 derive $M_V(HB) =
0.16 \cdot [Fe/H]  +1.02$) :
\begin{eqnarray}
M_V(HB) &=& 0.30 \cdot [Fe/H] + 0.94 \ ({\rm S\&T})\\
M_V(HB) &=& 0.20 \cdot [Fe/H] + 1.0 \ \ \ ({\rm Harris})
\end{eqnarray}

For deriving the Milky Way GCLF in the same way as the Fornax data, we used 
 the compilation of Harris (1994) \nocite{Harris:1994a} which contains the V magnitudes of all 
142 known clusters. The I magnitudes were calculated using the integrated color indices and 
a reddening correction (Dean et al. 1978)\nocite{Dean:1978}: 
\begin{equation}
M_{I,0}=M_{V,0}-((V-I)-1.35 \cdot E_{B-V})
\end{equation}
We made no restrictions concerning galactocentric distance and reddening. 
We fitted the GCLFs in V and in I after binning in 0.5 mag. The dispersion
$\sigma_{V}=1.2$ mag and $\sigma_{I}=1.1$ mag (for Gaussians) were determined by a free fit.
Since the turnover magnitudes are practically independent of whether a Gaussian or
a t5-function is fitted, Table \ref{milkyway} only contains the results of the Gaussian 
representation (this is of course not the case, if only the
ascendent branch of the GCLF or part of it is visible).

Corresponding to Table \ref{milkyway}, the distance moduli to the Fornax cluster
in V are $31.20 \pm 0.13$ and $30.89 \pm 0.13$ in I, using the RR-Lyrae scale of Harris et al. (1991). 
Following Burstein \& Heiles (1984)
\nocite{Burstein:1984} we adopted zero absorption.
In the next section, we argue that the smaller modulus
in I may be understood as a metallicity effect.   


 \begin{table}[t]
\caption[]{The absolute turnover magnitudes in V and I for the galactic GCLF.
 Different  relations for the metallicity dependence of the brightness of
 horizontal branch stars of Sandage $\&$ Tammann (1995) and Harris et al. (1991) were used.
The data have been binned in 0.5 mag bins.}
\begin{tabular}{ccc}
\hline
 $V_{0}$ & $I_{0}$ & calibration of\\
 $\sigma =1.2$ mag& $\sigma =1.1$ mag & RR-Lyrae stars\\
\hline
-7.40 $\pm 0.12$ & -8.53 $\pm 0.10$ & Harris et al. (1991) \nocite{Harris:1991} \\
-7.68 $\pm 0.12$ & -8.73 $\pm 0.10$ & Sandage \& Tammann (1995) \nocite{Sandage:1995}\\
\hline
\end{tabular}
\label{milkyway}
\end{table} 

\section{Metallicity corrections}

To what degree is the turnover of GCLFs universal? In particular, do the GCSs of
 spiral and elliptical galaxies have the same turnover? It is a
remarkable fact that the {\it mass functions} of globular clusters in galaxies which are so different
in their properties as the Milky Way and M87, exhibit the same mass function slope,
which may be regarded as the underlying property for a universal turnover (Harris \& Pudritz 1994
\nocite{Harris:1994}, McLaughlin \& Pudritz 1995\nocite{McLaughlin:1995}). 
If this is so, then one should expect a metallicity dependence in the turnover
since at a given mass, metal-poor globular clusters are brighter than metal-rich
clusters. This behaviour has been studied by Ashman et al. (1995) \nocite{Ashman:1995} assuming
a universal globular cluster mass function and using the (M/L)-[Fe/H] relationships given
by Worthey (1994) \nocite{Worthey:1994}.

In this sense, we also expect the GCS of spirals to be more metal-poor than those 
of ellipticals, because the metal-rich bulges and their associated GC populations are 
dominating, while in spirals like the Milky Way, the GC population of the
bulge comprises only a handful of clusters (e.g. Richtler 1995\nocite{Richtler:1995}). 
If this
is the case then we overestimate the distance modulus of a metal-rich GCS by
calibrating it through the galactic system. The larger distance modulus in V may
be understood in this way. To correct for it, we had to know the difference
in the mean metallicity between the galactic system and the Fornax GCSs.
Unfortunately, we only have the integral (V-I)-colour, which is no sensitive metallicity 
indicator. Adopting zero reddening towards the Fornax cluster and using 
the colour-metallicity relation of Couture (1990) \nocite{Couture:1990}
\begin{eqnarray}
(V-I)_{0}&=&0.2[Fe/H]+1.2
\end{eqnarray}
we calculated the shifts in the turnover on the basis of 
Ashman et al. (1995) with respect to the Milky Way for each
galaxy (see Table \ref{vi}). On average we get a mean Fornax turnover $<V_{TO}>=23.67 \pm 0.06$ 
in V and \mbox{$<I_{TO}>=23.32 \pm 0.05$} in I. The resulting distance moduli are $31.07 \pm 0.13$ in V 
and $30.85 \pm 0.11$ in I (using the RR-Lyrae scale of Harris et al. 1991).  

\section{The relative distance to Virgo}
The direct comparison of the turnover of GCLFs of elliptical galaxies only avoids
all the uncertainties in the predicted influence of metallicity and the difficulties
in its determination. A comparison of the turnover of galaxies in the Fornax
and the Virgo cluster therefore should yield an accurate relative distance.
The GCLF of M87 recently has been observed with the 
Hubble Space Telescope (Whitmore et al. 1995 \nocite{Whitmore:1995a}, Elson \& Santiago
1995\nocite{Elson:1995}). The deep photometry of Whitmore et al. reached
more than two magnitudes beyond the turnover of $V^{M87}_{TO}=23.72 \pm 0.06 $ mag. A galactic
extinction of $A_{V}=0.067 \pm 0.04$ mag towards M87 has been adopted.
Because of its position in the core of the Virgo cluster, M87 provides a good reference for
the relative distance modulus. 
Using the mean
Fornax turnover of section 4, we derive 
\begin{equation}
(m-M)_{Fornax} - (m-M)_{M87} = 0.08 \pm 0.09 ~\mbox{mag}
\end{equation}
in good agreement with the recent determination \mbox{$(m-M)_{Fornax} - (m-M)_{Virgo} = -0.06 \pm 
0.15$ mag}
 of Bureau et al. (1996) using the I-band Tully-Fisher relation.   
{\it Basically there is no difference in the distance of the Fornax cluster and of the Virgo 
cluster. }

        \begin{table}[t]
           \caption[]{Colours, predicted shifts and corrected turnover of the GCLFs according to 
Ashman et al. (1995). The metallicities were computed using equation (4).}
            \begin{tabular}{ccccccc}
            \hline
            galaxy & V-I & [Fe/H] & $\Delta V$ & $\Delta I$& $V^{*}_{0}$ & $I^{*}_{0}$ \\
            \hline
            NGC 1374 & 1.08 & -0.64 & 0.21 & 0.09 & 23.31 & 22.51\\
            NGC 1379 & 1.17 & -0.19 & 0.40 & 0.15 & 23.28 & 22.39\\
            NGC 1427 & 1.02 & -0.94 & 0.10 & 0.05 & 23.73 & 22.26\\
            NGC 1399 & 1.01 & -1.39 & 0.09 & 0.04 & 23.81 & 22.29\\
            \hline
            \end{tabular}
          \label{vi}
         \end{table}


\begin{acknowledgements}
  This research was partially supported by the DAAD and Deutsche Forschungsgemeinschaft 
(Ri 418/5-1) and by FONDECYT Grants No. 1930572 (to HQ) and 1930570 (to LI). 
Furthermore, HQ acknowledges the award of a Presidential Chair in Science.
\end{acknowledgements}

\end{document}